\documentclass[11pt,a4paper]{article}
\pdfoutput=1
\usepackage{jcappub}
\usepackage[utf8]{inputenc}
\usepackage[normalem]{ulem}
\usepackage{tikz}
\usepackage{pgf}
\usetikzlibrary{arrows,shapes}
\usepackage{amsfonts}
\usepackage{fullpage}

\usepackage{enumerate}

\newcommand{\be}{\nopagebreak[3]\begin{equation}}
\newcommand{\ee}{\end{equation}}
\newcommand{\bfig}{\nopagebreak[3]\begin{figure}}
\newcommand{\efig}{\end{figure}}
\newcommand{\ba}{\nopagebreak[3]\begin{eqnarray}}
\newcommand{\ea}{\end{eqnarray}}

\newcommand{\bmult}{\nopagebreak[3]\begin{multline}}
\newcommand{\emult}{\end{multline}}
\newcommand{\Fref}[1]{Fig.\,\ref{#1}}
\newcommand{\fref}[1]{Fig.\,\ref{#1}}
\newcommand{\eref}[1]{eq.\,(\ref{#1})}

\newcommand{\grad}{\nabla}
\newcommand{\h}{\hat}
\def\etai{\eta_{\rm ini}}

\def\f{\frac}

\def\ns{n_{\rm s}}
\def\nt{n_{\rm t}}
\def\as{A_{\rm s}}

\def\ig{\includegraphics}

\def\omm{\Omega_{\rm m}}
\def\omr{\Omega_{\rm r}}
\def\oml{\Omega_{\rm \Lambda}}
\def\omk{\Omega_{\rm k}}
\def\lcdm{$\Lambda{\rm CDM}~$}

\def\efolds{{\it e}-folds~}

\def\camb{\texttt{CAMB}~}

\def\V0{\mathcal{V}_o}

\def\celltt{C^{\rm TT}_\ell}
\def\cellbb{C^{\rm BB}_\ell}

\def\planck{{\it Planck} }

\def\mpc{{\rm Mpc^{-1}}}
\def\mpl{m_{\rm Pl}}
\def\s3{\mathbb S^3}

\newcommand{\Pt}{\mathcal{P}_t}

\begin{document}
\title{Tensor perturbations during inflation in a spatially closed Universe}

\author{B\'eatrice Bonga,}\emailAdd{bpb165@psu.edu}

\author{Brajesh Gupt,}\emailAdd{bgupt@gravity.psu.edu}

\author{Nelson Yokomizo}\emailAdd{yokomizo@gravity.psu.edu}

\affiliation{
Institute for Gravitation and the Cosmos \& Physics Department, The Pennsylvania State University, University Park, PA 16802 U.S.A.
}
\abstract{
In a recent paper \cite{bgy1}, we studied the evolution of the background geometry 
and scalar perturbations in an inflationary, spatially closed Friedmann-Lema\^itre-Robertson-Walker (FLRW) model having constant positive spatial curvature and 
spatial topology $\s3$. Due to the spatial curvature, the early 
phase of slow-roll inflation is modified, leading to suppression of power in the scalar 
power spectrum at large angular scales. In this paper, we extend the analysis to include 
tensor perturbations. We find that, similarly to the scalar perturbations, the tensor 
power spectrum also shows suppression for long wavelength modes. The correction to 
the tensor spectrum is limited to the very long wavelength modes, therefore the 
resulting observable CMB B-mode polarization spectrum remains practically the same as in 
the standard scenario with flat spatial sections. However, since both the tensor and
scalar power spectra are modified, there are scale dependent corrections to the
tensor-to-scalar ratio that leads to violation of the standard slow-roll consistency
relation.
}
\maketitle

\section{Introduction}
\label{s1}
Inflation is the leading paradigm of the early Universe, according to which  
the tiny temperature fluctuations observed in the cosmic microwave background (CMB)
originate from quantum vacuum fluctuations at very early times 
\cite{guth,albrecht,linde,lindechaotic,liddlelyth,mukhanov,weinberg}.
One of the compelling features of the inflationary scenario is that it leads to an
almost flat, homogeneous and isotropic Universe today starting from generic 
initial conditions. In particular, due to an exponential increases in the scale factor
of the Universe, inflation dilutes away any effects of spatial
curvature and as a result the spacetime is extremely well approximated by a spatially flat FLRW
model at the end of inflation. Conversely, even though the spatial curvature 
effects are subdominant today, they could have been important in the early phases of
inflation and in the pre-inflationary era. In addition, the equations of motion for
cosmological perturbations include new curvature terms that become important at these times. 
In this scenario, modes of metric perturbations exiting the Hubble horizon at these 
early times can carry imprints of spatial curvature in their power spectrum. 

Recent cosmological observations determine the density parameter due to the spatial
curvature to be $\omk=-0.005_{-0.017}^{+0.016}$ from \planck data alone and
$\omk=0.000_{-0.005}^{+0.005}$ by combining \planck measurements 
with BAO data \cite{Ade:2015xua}. While these values for $\omk$ are compatible with 
a flat FLRW model, they allow the possibility that our Universe is
spatially closed with spatial topology of a 3-sphere ($\mathbb S^3$). For
$\omk=-0.005$, the physical radius of this 3-sphere today is 
approximately $64~{\rm Gpc}$, which is $4.5$ times 
the physical radius of the CMB sphere. This indicates that the
effects of the spatial curvature, if present, would be most prominent for 
length scales similar to or larger than that of the CMB sphere.

The presence of spatial curvature can affect observations in two 
ways: 
First, by affecting the background spacetime evolution in the post-inflationary era.
This has been studied in detail by taking into account the
consequences of spatial curvature on the Friedmann and Boltzmann
equations in radiation and matter dominated era \cite{Lewis:1999bs}. In this context,
for a specified primordial power spectrum, the effect of $\s3$ spatial topology
during post-inflationary era on the CMB is numerically computed using the publically 
available code \camb \cite{Lewis:1999bs,camb_notes}. Second, the spatial topology also 
introduces corrections to the primordial power spectrum of the scalar and tensor
perturbations generated during inflation. This is the 
focus of the current and the accompanying paper \cite{bgy1}. Corrections to 
the primordial power spectrum of scalar perturbation due to $\s3$ spatial topology
have been previously studied using approximate methods 
\cite{Uzan:2003nk,Ellis:2001ym,Ellis:2001yn,Efstathiou:2003hk,Lasenby:2003ur,Luminet:2003dx}.
In the accompanying paper \cite{bgy1}, we studied the evolution of the background 
geometry and gauge-invariant scalar perturbations in a spatially closed FLRW Universe 
where inflation is driven by a single scalar field in the presence of an inflationary 
potential (taken to be either quadratic or Starobinsky potential). We found that
effects of non-vanishing spatial curvature become important during the early phases of 
inflation ($\sim60$ \efolds before the end of inflation). Therefore, the long wavelength 
modes which exit the Hubble radius at those times can carry imprints of spatial 
curvature in their power spectrum. For both potentials, the resulting scalar power 
spectrum is different from the almost scale invariant power spectrum obtained in a 
spatially flat inflationary FLRW spacetime for long wavelength modes. In particular, 
the power spectrum shows oscillatory behavior that is most prominent for the 
long wavelength modes. In addition, there is suppression of power at such scales as 
compared to the flat model. For $\omk=-0.005$, the power deficit in the primordial 
power spectrum leads to suppression of power in the temperature anisotropy spectrum 
$\celltt$ at $\ell<20$ \cite{bgy1}. For small wavelength modes, the power spectrum 
approaches the nearly scale invariant power spectrum and the resulting $\celltt$ agrees 
extremely well with that in the flat model for $\ell \gtrsim 20$. 

In every model of inflation, in addition to scalar fluctuations, the quantum vacuum 
fluctuations in the early Universe
are also expected to generate tensor metric perturbations \cite{Starobinsky:1979ty},
which contribute to the B-mode polarization in the CMB.  One of the main goals of the
current and upcoming CMB experiments is the measurement and characterization of these 
primordial B-mode signals \cite{spt,bicep,polarbear,class}. Furthermore, the tensor 
perturbations generated during inflation give rise to primordial gravitational waves 
which are expected to contribute to the stochastic gravitational wave background 
at very low frequencies. Such low-frequency stochastic gravitational wave background 
can potentially be observed by space-based gravitational wave detectors such as eLISA
\cite{elisascience,elisagw}. Since the primordial gravitational waves propagate through 
the Universe without being attenuated by the intermediate matter fields, they  
provide a direct window on the physics of the inflationary \cite{Zaldarriaga:1998ar} 
as well as pre-inflationary era of the early Universe. 
Tensor perturbations have been studied in great detail in the context of
spatially flat and open FLRW models 
\cite{Mukhanov:1990me,Turner:1993vb,Lidsey:1995np,Baumann:2009ds}. The goal of this paper 
is to investigate the effect of spatial curvature on the spectrum of primordial tensor 
perturbations generated during inflation in a spatially closed inflationary FLRW
model as well as its effect on the B-mode polarization.

In this context we are led to the following questions: 
Scalar modes exiting the curvature radius during the early stages of inflation 
show power suppression. Is there similar suppression for
tensor modes as well? If so, what effects does it produce in the CMB B-mode
polarization signal and at what scales? 
One of the predictions of the single field slow-roll inflationary scenario is 
the so called slow-roll consistency relation that relates the tensor-to-scalar 
ratio $r$ with the tensor spectral index $\nt$ \cite{Mukhanov:1990me,Lidsey:1995np,Baumann:2009ds}. This 
slow-roll consistency relation was derived for a flat FLRW model with Bunch-Davies
initial conditions for the perturbations. Since the early slow-roll phase is modified in 
the closed model due to the presence of spatial curvature \cite{bgy1}, 
it is natural to ask if the 
tensor-to-scalar ratio is also modified and whether or not the slow-roll consistency 
relation still holds true.\footnote{The slow-roll consistency relation can also be
modified in the spatially flat FLRW model, for instance, by considering excited
initial states for the quantum perturbations \cite{aan3,LelloFastRoll}.} 
The potential deviations from the standard slow-roll consistency relation, 
if measured by future experiments, can then be potentially used to refine the 
constraints on the spatial curvature of the Universe. 

Our analysis shows that the tensor power spectrum is also suppressed at long
wavelength scales, which is qualitatively similar to that in the scalar power spectrum. 
However, power is suppressed by a few percent for tensor perturbations as opposed 
to that for scalar perturbations in which case the suppression is as large as $10\%$ 
for the largest observable mode (corresponding to $\ell=2$) for $\omk=-0.005$. 
Due to the small suppression, the resulting B-mode polarization spectrum 
$\cellbb$ remains practically unchanged by the presence of positive spatial curvature. 
Since the scalar and tensor spectra are modified differently, the tensor-to-scalar
ratio $r$ for a closed FLRW model is different from that for the flat model at long
wavelength modes. Specifically, there is enhancement in $r$ at those scales.
As a result, the slow-roll consistency relation is also modified for long wavelength modes.
As shown in \cite{bgy1}, short wavelength modes which exit the
curvature radius later during inflation are not affected by the presence of spatial
curvature. Therefore, $r$ and the slow-roll consistency relation for short wavelength
modes for the spatially closed FLRW model are the same as that for the flat model. 

The paper is organized as follows. In section \ref{s2.1}, we provide a brief review of
the evolution of the background spacetime for a spatially closed FLRW model. In section
\ref{s2.2}, we derive the  equation of motion for tensor perturbations and discuss the
choice of initial conditions. In section \ref{s3.1}, we study the numerical evolution of
tensor perturbations, compute the tensor primordial power spectrum, its impact on the 
B-mode polarization spectrum and compare the results with 
those for scalar modes. In section \ref{s3.2}, we discuss the
modifications in the tensor-to-scalar ratio and slow-roll consistency relation due to  
positive spatial curvature. We summarize the main results and discuss future outlook in
section \ref{s4}.

\section{Spatially closed FLRW background geometry and tensor perturbations}
\label{s2}
This section is composed of two subsections. In section \ref{s2.1}, we briefly
revisit the equations governing the background geometry of a spatially closed FLRW
model in the presence of an inflationary potential. In section \ref{s2.2}, we will 
discuss the linear tensor metric perturbations $h_{ab}$ off the closed FLRW geometry. 
We will obtain the equations of motion of $h_{ab}$ and discuss the choice of initial 
conditions.
\subsection{Background: dynamics and initial conditions}
\label{s2.1}
We consider a homogeneous, isotropic spatially curved FLRW model with topology 
$\mathbb{R} \times \mathbb{S}^3$ where $\mathbb{S}^3$ refers to the topology of the 
spatial sections. Choosing the scale factor today to be unity, i.e. $a_0=1$,
the value of the curvature radius of the $\mathbb S^3$ spatial section $r_o$
can be expressed in terms of the Hubble constant $H_0$ and the spatial curvature 
parameter $\omk$ as $r_o = 1/\sqrt{H_0^2 |\omk|}$. With this convention, the spacetime 
metric is 
\begin{equation}
\label{eq:metric}
ds^2 = - dt^2 + a^2(t)~r_o^2 \, \left( d\chi^2 + \sin^2\chi \, 
\left(d\theta^2 + \sin^2 \theta \, d \varphi^2 \right)
\right),
\end{equation}
where $a(t)$ is the dimensionless scale factor.\footnote{Note
that in \cite{bgy1} we used a scale factor which carried dimensions of length. In
this paper, we work with a dimensionless scale factor while $r_o$ carries the
dimension of length.} The unit 3-sphere on the spatial section $\s3$ is coordinatized by 
$(\chi, \theta, \phi)$ which are defined in the domain: $\chi, \theta \in \left[ 0 , \pi
\right]$; $\phi \in \left[0, 2 \pi \right]$. In conformal time, defined by $dt = a d\eta $, the metric reads:
\begin{equation}
ds^2 = a^2(\eta) \left[ - d \eta^2 + r_o^2 \, \left( d\chi^2 + \sin^2\chi \left(d\theta^2 + \sin^2 \theta \, d \varphi^2 \right) \right) \right] \, .
\label{eq:metric-conformal-time}
\end{equation}
The metric associated with the line element \eqref{eq:metric-conformal-time} can be
written as $g_{ab}=a^2(\eta) \mathring{g}_{ab}$, where $\mathring{g}_{ab}$ describes
a static metric with spherical spatial sections of radius $r_o$, that is, an Einstein Universe of radius $r_o$.

Given a matter field with energy density $\rho$, pressure $p$ and vanishing
anisotropic stress, the dynamics of the background spacetime is governed by the 
Friedmann and Raychaudhuri equations: 
\begin{eqnarray}
&H^2& = \frac{8 \pi G}{3} \rho - \f{1}{a^2r_o^2},
\label{eq:friedmann} 
\\
&\dot{H}& = - 4 \pi G \left( \rho + p \right) + \f{1}{a^2r_o^2},
\label{eq:raychaudhuri}
\end{eqnarray}
where $H=\dot{a}/a$ is the Hubble parameter and `dot' is used to represent the
derivative with respect to proper time. Here we are interested in studying the
dynamics of the background geometry during inflation in the presence of a single scalar 
field with a self-interacting potential $V(\phi)$. The energy density and pressure of
the scalar field are then given by:
\be
 \rho= \f{1}{2} \dot{\phi}^2 + V(\phi) \qquad {\rm and } \qquad p= \f{1}{2}
\dot{\phi}^2 - V(\phi). 
\ee
Conservation of the energy momentum tensor of the scalar field results in the
Klein-Gordon equation:
\be
  \ddot \phi + 3 H \dot\phi + \partial_\phi V(\phi) = 0.
\ee
For a single field inflationary model, recent \planck data favors the
Starobinsky potential over other potentials,
therefore in this paper we will restrict our analysis to the Starobinsky
potential:\footnote{Another motivation for using the Starobinsky potential is as follows. 
This potential naturally arises after conformally transforming a modified theory of gravity. 
Specifically, starting with a higher derivative Lagrangian of the 
form $R+\alpha R^2$, upon a conformal transformation, this theory can be written as the 
Einstein-Hilbert action and a scalar field with the potential \eref{eq:starobinsky}
\cite{Barrow:1988xh,Maeda:1988ab,Starobinsky:2001xq,DeFelice:2010aj,Ade:2015lrj}.} 
\be
V(\phi) = \f{3m^2}{32 \pi G} \left(1- e^{-\sqrt{\f{16\pi G}{3}} \phi} \right)^2. \, 
\label{eq:starobinsky}
\ee
Nonetheless, the qualitative features of the results presented here also 
hold for other inflationary potentials. Specifically, the result that the
power spectrum is modified for long wavelength can be attributed to the topology 
$\mathbb S^3$. However, the specifics of the modification and the scale at 
which the deviations from the scale invariant power spectrum appear depend on 
the interplay between the topology and the inflationary potential.

Usually, in the case of a spatially flat inflationary FLRW spacetime, the value of the 
mass parameter $m$ in \eref{eq:starobinsky} is determined using the amplitude of the
scalar power spectrum $\as$ and the associated spectral index $\ns$ at the comoving pivot 
scale $k_\star$ from observations in conjunction with Einstein's equations 
\cite{ag3,bg1,bg2,bgy1}. This
procedure also yields the values of $H$, $\phi$, $\dot\phi$ and $a$ 
(in the convention with $a_0=1$ today) at the time $t=t_\star$, i.e. when the
pivot mode $k_\star$ exits the Hubble radius during inflation. Here, we will
use the data from the \planck mission for which amplitude of the scalar power
spectrum $\as=(2.2065 \pm 0.0035) \times 10^{-9}$ and scalar spectral index $\ns=0.9645 \pm 0.0049$ 
at the pivot scale $k_\star=0.05~\mpc$. 
In principle, for a closed FLRW model, the values of various
parameters at $t=t_\star$ would be different from that in the flat model and one
would need to re-compute them. However, it turns out that at $t=t_\star$ the ratio of 
the total energy density due to the spatial 
curvature and that due to the inflaton field remains smaller
than $10^{-6}$ for the range of $\omk$ given by \planck measurements. Therefore, the
conditions at $t_\star$ are practically the same as those in the spatially flat FLRW
model \cite{bgy1}.  We obtain the following conditions at $t=t_\star$:\footnote{For more details concerning the initial data and the determination of
the mass parameter, see section 3 in \cite{bgy1}.}
\ba
a_\star &=& (2.11\pm 0.14) \times 10^{-53}, \qquad \quad \; \; \phi_\star=1.06 \pm 0.057 ~\mpl,
\nonumber \\ 
H_\star &=& (1.25\pm 0.32)\times 10^{-6}~\mpl^2, \qquad \dot{\phi}_\star = (-5.27
\pm 3.18)\times 10^{-9}\mpl^2,
\label{eq:staroini}
\ea
where subscript `$\star$' indicates that the quantities are evaluated at horizon exit time 
$t=t_\star$. The mass parameter is
\begin{equation}
m= (2.67 \pm 0.18) \times 10^{-6}~\mpl.
\end{equation} 
As shown in \fref{fig:horizon}, the effects of $\omk$ become 
important approximately 8 \efolds before $t_\star$.
\bfig
  \ig[width=0.7\textwidth]{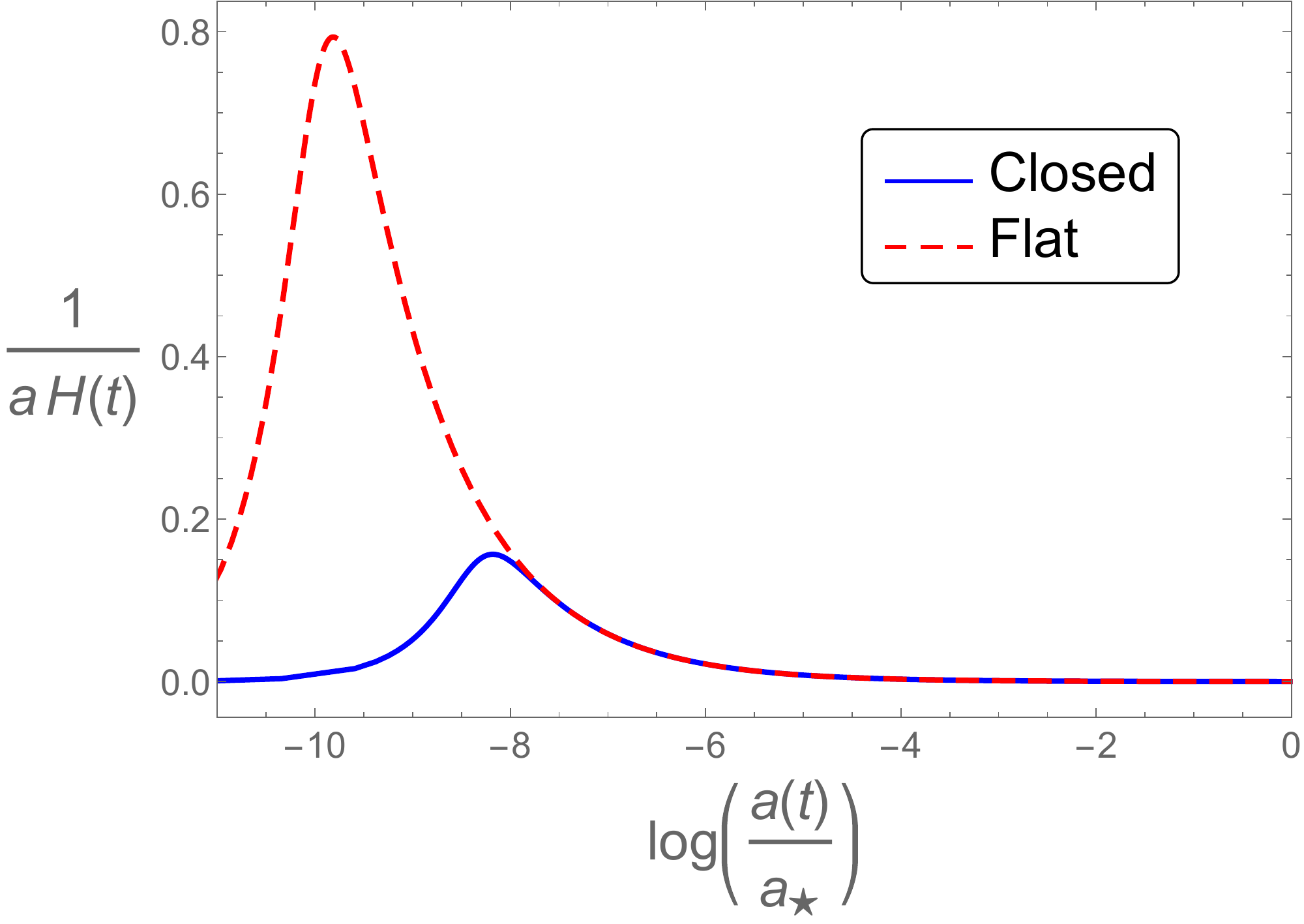}
  \caption{Comoving Hubble horizon $(a H)^{-1}$ plotted with respect to the number of
e-folds from the horizon exit time $t_\star$ of the pivot mode. The dashed (red)
and the solid (blue) curves correspond to spatially flat and closed models
respectively. It is evident that the spatial curvature starts becoming important
approximately 8 \efolds before $t_\star$. The onset of accelerated expansion marked
by $\ddot a=0$---which corresponds to the maximum of the curves---happens approximately 10 \efolds before $t_\star$ in the flat model
and apprximately 8 \efolds before $t_\star$ in the closed model.}
\label{fig:horizon}
\efig
Within the range of allowed conditions at $t_\star$, the condition that maximize the 
effect of spatial curvature corresponds to the situation for which the size of the comoving 
wavelength of the largest observable mode is comparable to the comoving 
Hubble horizon $(aH)^{-1}$ at the onset of inflation. 
This initial condition will be contrasted to the initial condition corresponding
to the best fit value of the parameters $\as$ and $\ns$ in the results in the 
next section. Specifically, for the `maximal effect' condition we have:
\begin{equation} 
\phi_\star= 1.02  ~\mpl, \qquad 
\dot{\phi}_\star = -6.69 \times 10^{-9}\mpl^2.
\label{eq:staroiniextreme}
\end{equation}

The evolution of the background spacetime after the end of inflation, which is marked by 
$\epsilon:=-\dot{H}/H^2=1$, is governed by the \lcdm model. We will assume instantaneous 
reheating and will use $\omm=0.31,~\omr=9.2\times10^{-5},~\oml=0.69$
as determined by {\it Planck}. For the spatial curvature, we work with 
$\omk=-0.005$ as a representative value to show plots and discuss numerical 
results.\footnote{We use $\omk=-0.005$ as a representative value because it is the best 
fit value determined from the data by \planck and lies at the 95\% confidence interval when one includes BAO data in addition to \planck data.}

\subsection{Tensor perturbations: dynamics and initial conditions}
\label{s2.2}
We introduce linear perturbations on the FLRW background with spherical spatial surfaces described in the previous section by splitting the metric into a background 
and a perturbation part. Specifically, we consider a one-parameter family of metrics
$g_{ab} (\varepsilon)$:
\begin{equation}
g_{ab}(\varepsilon) = a^2(\eta) \left( \mathring{g}_{ab} + \varepsilon h_{ab} \right) \, ,
\end{equation}
where $a^2 \mathring{g}_{ab}$ is the background metric \eqref{eq:metric-conformal-time}, $h_{ab}$ is a linear perturbation and $\varepsilon$ is a smallness parameter. Indices of the perturbation $h_{ab}$ are raised and lowered with $\mathring{g}_{ab}$ and its inverse $\mathring{g}^{ab}$. Since we are interested only in the tensor components of $h_{ab}$, 
we can focus on its transverse traceless part $h_{ab}^{TT}$, that is,
\begin{align*}
h_{ab}^{TT} \eta^b = 0, \qquad \qquad \mathring{\grad}^a h_{ab}^{TT} =0 \qquad {\rm and} \qquad \mathring{g}^{ab} h_{ab}^{TT} =0,
\end{align*}
with $\eta^a$ the globally time-like conformal Killing vector field of 
the background metric $\mathring{g}^{ab}$ and 
$\mathring{\grad}_a$ the derivative operator compatible with $\mathring{g}_{ab}$.\footnote{Here, abstract index notation is used \cite{Wald:1984rg}.}
The equations of motion of the tensor perturbations are then obtained by linearizing
Einstein's equations.\footnote{The \eref{eq:eomtensor} can also be obtained from the quadratic action for the
perturbations $h_{ij}$ given as:
\be
S^{(2)} = \frac{1}{64\pi G} \int d^4x~a^2~r_o^3 \sin^2 \chi \sin \theta \left( h'^{ij} h_{ij}' -  \mathring{\nabla}_k h_{ij} \mathring{\nabla}^k h^{ij} - \frac{2}{r_o^2} h^{ij} h_{ij} \right) \, .
\ee
} For the nonvanishing spatial components, we find:
\begin{equation}
\label{eq:eomtensor}
\mathring{\grad}^2 h_{ij}^{TT} - 2 \frac{a'}{a} h_{ij}^{' \, TT} - \f{2}{r_o^2} h_{ij}^{TT} = 0 \, ,
\end{equation}
where the `prime' denotes the partial derivative with respect to conformal time $\eta$
and $i,j=1,2,3$ denote spatial indices. Note that although we have used a specific background, this equation is true for any 
FLRW model with matter field(s) having vanishing anisotropic stress.

Recall that in the case of flat spatial sections, instead of solving the equations of
motion in real space, one expands the perturbations into their Fourier modes. The
equations of motion then become an ordinary differential equation for each Fourier
mode. Similarly, in the spatially closed model considered here, it is convenient to
expand the perturbations in a basis spanned by the tensor hyperspherical harmonics 
$\mathcal{Q}^{nlm,s}_{ij}(\chi,\theta,\phi)$ \cite{Gerlach:1978gy}. Let $\gamma_{ij}$ be the metric of the three-sphere $\s3$ with radius $r_o$ and $D_i$ the corresponding covariant derivative. These hyperspherical harmonics are tensor eigenfunctions of the Laplacian operator $D^2$:
\begin{equation}
D^2 \mathcal{Q}^{nlm,s}_{ij}(\chi,\theta,\phi)=-\f{(n^2-3)}{r_o^2} \mathcal{Q}^{nlm,s}_{ij}(\chi,\theta,\phi) \, ,
\end{equation}
satisfying
\begin{align*}
\gamma^{ij} \, \mathcal{Q}^{nlm,s}_{ij} =0 \, ,
\qquad
D^i \mathcal{Q}^{nlm,s}_{ij} =0 \, ,
\end{align*}
and form an orthonormal basis in the space of transverse traceless tensor fields on $\s3$. They are labelled by integers in the intervals: $n\geq 3$, $l \in[0, n-1]$,
$m\in [-l, l]$. The polarization index $s$ assumes values $s=0,1$, which distinguish between even and odd harmonics with respect to the parity operator
\begin{equation}
P_2(\chi,\theta,\varphi) = (\chi,\pi-\theta,\pi+\phi) \, ,
\end{equation}
i.e., $\mathcal{Q}^{nlm,s}_{ij}\left(P_2(\chi,\theta,\varphi)\right)=(-1)^{l+s}
\mathcal{Q}^{nlm,s}_{ij}(\chi,\theta,\varphi)$. For an explicit expression of these
tensor hyperspherical harmonics and their derivation, see \cite{Gerlach:1978gy,
Challinor:2000as}. The factor $(n^2-3)/r_o^2$ can be interpreted as the curved
version of the wavenumber squared ($k^2$) by analogy with the flat model for which
the eigenfunctions of the flat space Laplacian have eigenvalue $-k^2$ 
\cite{Lewis:1999bs}.\footnote{For scalar modes, the eigenvalue of the Laplacian on
the scalar hyperspherical harmonics is $-(n^2-1)$ and thus for scalar modes $n^2-1$,
instead of $n^2-3$, is the curved space version of $k^2$.}

We expand the tensor perturbations in terms of the tensor hyperspherical harmonics as
follows: 
\be
h_{ij}^{TT}(\eta,\chi,\theta,\phi) = \sum_s \sum_{nlm} h^s_{nlm}(\eta)
\mathcal{Q}^{nlm,s}_{ij}(\chi,\theta,\phi),
\label{eq:modes}
\ee
All dynamical information of $h_{ab}^{TT}$ is now 
encoded in the amplitudes $h^s_{nlm}(\eta)$ which are purely functions of time.
Substituting this decomposition into \eref{eq:eomtensor} yields the following equation of 
motion for $h^s_{nlm}$: 
\begin{equation}
h^{s~''}_{nlm} + 2 \frac{a'}{a} h^{s~'}_{nlm} + \left(\frac{n^2 -3}{r_o^2} +
\f{2}{r_o^2}  \right) h^{s}_{nlm} = 0 \, .
\label{eq:eomconformaltime}
\end{equation}
Introducing the variable $\mu^s_{nlm}(\eta)=a(\eta) h^s_{nlm}(\eta)$, we obtain:
\begin{equation}
\mu^{s~''}_{nlm} + \left( \frac{n^2-1}{r_o^2} - \frac{a^{''}}{a} \right) \mu^s_{nlm} = 0 \, .
\label{eq:mueta}
\end{equation}

The field $\h{\mu}^s_{nlm}(\eta)$ behaves like a scalar field and can be written in the
standard Fock representation as follows:
\be
  \h\mu^s_{nlm}(\eta) = \h{A}_{nlm}^s~e_{nlm}^s(\eta) +
{\h{A}_{nlm}}^{s~\dagger}~{e^{s\,\star}_{nlm}}(\eta).
\ee
Here, $\h{A}^{s~\dagger}_{nlm}$ and $\h{A}^s_{nlm}$ are the creation and annihilation operators and 
the mode functions $e_{nlm}^s(\eta)$ form the positive frequency basis. Since the two
polarizations denoted by the superscript `$s$' evolve independently, in the following
we will drop this superscript for notational clarity. The mode functions then
satisfy the following equations of motion:
\be
e^{''}_{nlm}  + \left(\frac{n^2 -1}{r_o^2} - \frac{a^{''}}{a}\right) e_{nlm} = 0 \, .
\ee 
The normalization condition for $e_{nlm}$ is 
\be
 e_{nlm} e^{\star\prime}_{nlm} - e^\prime_{nlm} e^\star_{nlm} = i~.
\ee
The choice
of the vacuum state $|0\rangle$ annihilated by $\h{A}_{nlm}$ is associated with the choice
of the positive frequency basis $e_{nlm}$. Recall that in a static FLRW Universe 
(i.e. a Universe for which $a(t)$ is constant), the vacuum state can be uniquely 
selected by requiring it to be regular and invariant under all spatial symmetries and 
time-translations. This state then automatically minimizes the Hamiltonian for tensor
perturbations.
For a dynamical, time-dependent background such as the FLRW model considered here, 
there is no such state that minimizes the Hamiltonian at all times. In order to
compare our results with those for the spatially flat scenario, we work with an 
\textit{instantaneous} vacuum state. This state is defined by the basis function $e_{nlm}$ that minimizes the
Hamiltonian for the perturbations at a given time $\etai$ \cite{Mukhanov:2007zz,Fulling:1989nb}.
The basis mode functions for this instantaneous vacuum are:
\ba
e_{nlm}(\etai) & =& \frac{1}{\sqrt{2\omega_n \, r_o}} \, , \nonumber \\
{e}^\prime_{nlm}(\etai) & =& -i \sqrt{\frac{\omega_n}{2 \, r_o}} \, 
\label{eq:inipert}
\ea
with $\omega_n = \sqrt{\frac{n^2-1}{r_o^2}-\f{a^{''}}{a}}$ \cite{bgy1}. The initial 
conditions are given at the onset of inflation defined by $\ddot a=0$.

Since the Hamiltonian is minimized at an instant of time, the vacuum state 
chosen here depends on the time at which the initial conditions are specified. 
Nevertheless, the time dependence of the instantaneous vacuum state chosen 
is weak for all observable modes. This is because for the initial conditions 
for the background spacetime all observable modes are within the curvature 
radius at the initial time $\etai$. As a result, the initial state for the 
perturbations is nearly static close to the onset of inflation. In the numerical 
evolution of the modes, we explicitly verified this.

\vskip0.5cm
\noindent{\it Remarks:} 
\\
(i) In the standard inflationary scenario with spatially flat sections, the initial
states for quantum perturbations representing scalar and tensor perturbations are
often chosen to be the Bunch-Davies vacuum state at a time when all obervable modes are
inside the curvature radius and the spacetime can be well approximated by a de Sitter
spacetime. From \fref{fig:horizon} it is clear that, 8 \efolds before
$t_\star$ when all observable modes can safely be assumed to be within the Hubble
horizon, the flat model is still well approximated by a quasi-de Sitter spacetime.
Therefore, it is safe to assume that all modes are in the Bunch-Davies state 8 or 
more \efolds before $t_\star$. This is no longer true in the closed model. As shown in 
\fref{fig:horizon}, the spatial curvature becomes important at those times and the 
spacetime is no longer quasi-de Sitter. Therefore, it is not justifiable to take the 
quantum perturbations in the Bunch-Davies vacuum state. Moreover, the spacetime is
dynamical and the symmetries of the spacetime are not enough to select a unique
vacuum state unlike in the de Sitter spacetime. There remains a large freedom in the
choice of initial state. As discussed above, in this paper, we choose an
instantaneous vacuum state at the time when $\ddot{a}=0$ for the quantum perturbations.
This choice is adopted for both the flat and closed models for the comparison of the
predicted power spectra, in order to isolate effects due to the presence of spatial
curvature from those related to the choice of the initial vacuum. 
%
%
%
%
\\
\\
(ii) For observable modes, the static vacuum chosen here also satisfies the WKB 
approximation. If we had chosen the instantaneous state at a
time before $\ddot a=0$ when the slow-roll and WKB approximations are both violated, 
then the power spectrum of the long wavelength modes would change while that for the 
short, ultraviolet modes would remain the same as that in the flat model.
\\
\\
(iii) In this paper, we have taken the topology of the homogeneous, spatial slices to be
$\mathbb{S}^3$. This is the simplest topology compatible with a constant positive spatial curvature in three-dimensions, 
and provides a natural framework for the analysis of effects due to the deviation
from spatially flat geometry. The local 
geometry does not fix the global topology uniquely, however, and our results can be extended to various other topologies 
compatible with a locally spherical geometry. Signatures of a nontrivial topology in the temperature and polarization CMB 
spectra have been heavily studied for flat and open spatial sections (see for
instance, \cite{cornish,Fabre:2013wia}), and a family of spherical spaces --
specifically lens and prism spaces -- has been studied in \cite{Uzan:2003ea,Riazuelo:2002ct} for which the mathematical and numerical groundwork was laid in \cite{Lehoucq:2002wy,Lehoucq:2003}. 
The allowed global metrics compatible with a locally spherically symmetric geometry
were classified in \cite{seifert}, and have the generic form $\mathbb{S}^3/\Gamma$,
where $\Gamma$ is a subgroup of isometries of $\s3$ acting freely and discontinuously.
The topology is modified through the identification of points on 
$\mathbb{S}^3$ related by the action of elements of $\Gamma$, and an infinite number of topologies can be constructed 
in this way, falling in a small number of classes \cite{luminet}. For example, take the simplest nontrivial topology 
one can build this way: the projective space, $RP^3=\mathbb{S}^3/\mathbb{Z}_2$, which is obtained by identifying
antipodal points. Given a point on the sphere with coordinates $(\chi, \theta,\phi)$, the cosmological parity operator 
${P}_3$ relates antipodal points:
\begin{equation}
{P}_3 =
\begin{cases}
\chi \to \pi - \chi, \\
\theta \to \pi - \theta, \\
\phi \to \pi + \phi \,.
\end{cases}
\end{equation}
Since antipodal points are identified for the construction of $RP^3$, only modes with
the same value at antipodal points contribute. In other words, only those
hyperspherical harmonics with $n$ odd are present in $RP^3$. Since at the linearized
level all modes evolve independently, this does not alter their dynamics and hence
the resulting primordial power spectrum is not affected for the odd modes while all
even modes are forced to vanish. If one
goes beyond the linearized level, modes will couple and this is no longer true. 
\\
\\
(iv) As a side observation, for a spatially flat Universe, taking the massless limit in 
the equation of motion for the scalar modes yields the equation of motion for the 
tensor modes. This is not true for a Universe with closed spatial topology.

\section{Results}
\label{s3}
This section is divided into two subsections. In the first, we will calculate the
tensor power spectrum at the end of inflation and the resulting B-mode polarization
spectrum in the CMB. In the second subsection, we will discuss the violation of the
slow-roll consistency relation due to the spatial curvature. 
Recall from section \ref{s1} that the presence of the nonvanishing spatial curvature can
affect observations in two distinct ways. First, by introducing corrections to the
Friedmann equations and equations of motion of the perturbations during inflation.
As shown in \cite{bgy1}, the energy density due to spatial curvature becomes
important during the early stages of inflation. As a result, although the background
geometry of a spatially closed inflationary FLRW spacetime at the end of inflation
is indistinguishable from a spatially flat FLRW spacetime, the modes of quantum fields 
representing scalar and tensor metric perturbations, which exit the curvature
radius during early times can carry imprints of spatial curvature in their power spectrum. 
The second effect of the spatial curvature is due to the
fact that in $\s3$ topology the wavenumbers of the quantum perturbations are discrete
rather than a continuum as in the flat model. This effect becomes important in the
post-inflationary era in the computation of the anisotropy spectrum in the CMB
by solving the Boltzmann equations starting from an initial condition at the end of
inflation. Only a finite number of discrete modes contribute to the temperature anisotropy 
observed in the CMB and the final expression of $C_\ell$ is a discrete sum over the modes 
rather than an integral. As shown in \cite{bgy1} the aggregate effect for the scalar modes was  
suppression in the temperature anisotropy spectrum $\celltt$ in the CMB at multipoles 
$\ell<20$. As we will see in this section, 
the tensor modes also show deficit of power at those scales, but the percentage 
suppression is much smaller than that for scalar modes for the same background 
inflationary solution. For the discussion of the results, we will use the initial
conditions for the background perturbations as described in the previous sections
with $\omk=-0.005$.

\subsection{Primordial power spectrum and B-mode polarization spectrum}
\label{s3.1}

In order to study the power spectrum of the quantum perturbations we numerically
solve the equation of motion subject to the initial conditions discussed in
section \ref{s2}. We work with the canonical representation for the scalar ($v$) and 
tensor ($\mu$) perturbations which are governed by the following equations of
motion.\footnote{The scalar modes $v$ are related to the variable
$q$ used in \cite{bgy1} via $v=aq$.}
For scalar modes:
\be
  \ddot v_n(t) + b^{(S)}(n, t)~\dot v_n(t) + c^{(S)}(n, t)~v_n(t) = 0, 
\ee
and for tensor modes
\be
  \ddot \mu_n(t) + b^{(T)}(n, t)~\dot \mu_n(t) + c^{(T)}(n, t)~\mu_n(t) = 0,
\ee
where
\begin{multline}
b^{(S)}(n, t)  = H + \f{32 \pi  G a^3 \dot{a} \dot{\phi} V'(\phi) 
              +48 \pi  G a^2 \dot{a}^2 \dot{\phi}^2   
         -8 \pi  G a^2 \dot{\phi}^2 \left(8 \pi  G a^2 \left(\dot{\phi}^2-2 V \right)
               +2\right)}{2 a \dot{a} \left(2(n^2-4)\dot{a}^2 
              + 8 \pi G a^2 \dot{\phi}^2 \right)}~,
\end{multline}
\begin{multline}
 c^{(S)}(n, t)  =  \f{8\pi G}{ a^2 \dot{a}^2 \left(2 \left(n^2-4\right) 
                \dot{a}^2+8 \pi G  a^2 \dot{\phi}^2\right)} \\ 
         \Bigg[ \f{\dot{a}^4 (n^2-4) \left( n^2-1 + a^2 V'' \right)}{4\pi G}+ \left(4
n^2-7\right) a^3 \dot{a}^3 \dot{\phi} V' -\pi G \frac{n^2-1}{n^2-4} a^4\dot{\phi}^4 
         \left[8 \pi  G a^2 \left(\dot{\phi}^2+2 V\right)-6\right] \\
      +   (n^2-1) a^2 \dot{a}^2  \Bigg(  
            -6 \pi  G  \frac{n^2-5}{n^2-4} a^2\dot{\phi}^4+ 
             4 \pi G a^2 \dot{\phi}^2 V+\frac{3}{2} \dot{\phi}^2 + \frac{9}{2}
\dot{a}^2 \dot{\phi}^2 \Bigg) \\
                      + a^3 \dot{a} \left[ a \dot{a} \dot{\phi}^2 V''+2 a \dot{a}
V'^2+ 4 \pi G a^2 \dot{\phi} V' \left(\dot{\phi}^2+2 V\right)- \dot{\phi} V'  \right]
\Bigg]
- H~b^{(S)}(n, t) - \f{\ddot a}{a}~,
\end{multline}
\be
b^{(T)}(n, t)  = H~,
\ee
and 
\be
c^{(T)}(n, t)  = \f{n^2-1}{r_o^2~ a^2} - H^2 - \f{\ddot a}{a}.
\ee
Here, we have written the equations in cosmic time which is what is used for the
numerical evolution of the modes.  For notational convenience we have dropped the
sub- and superscripts denoting spherical wavenumbers and polarization for the tensor 
modes. 

\bfig
\ig[width=0.47\textwidth]{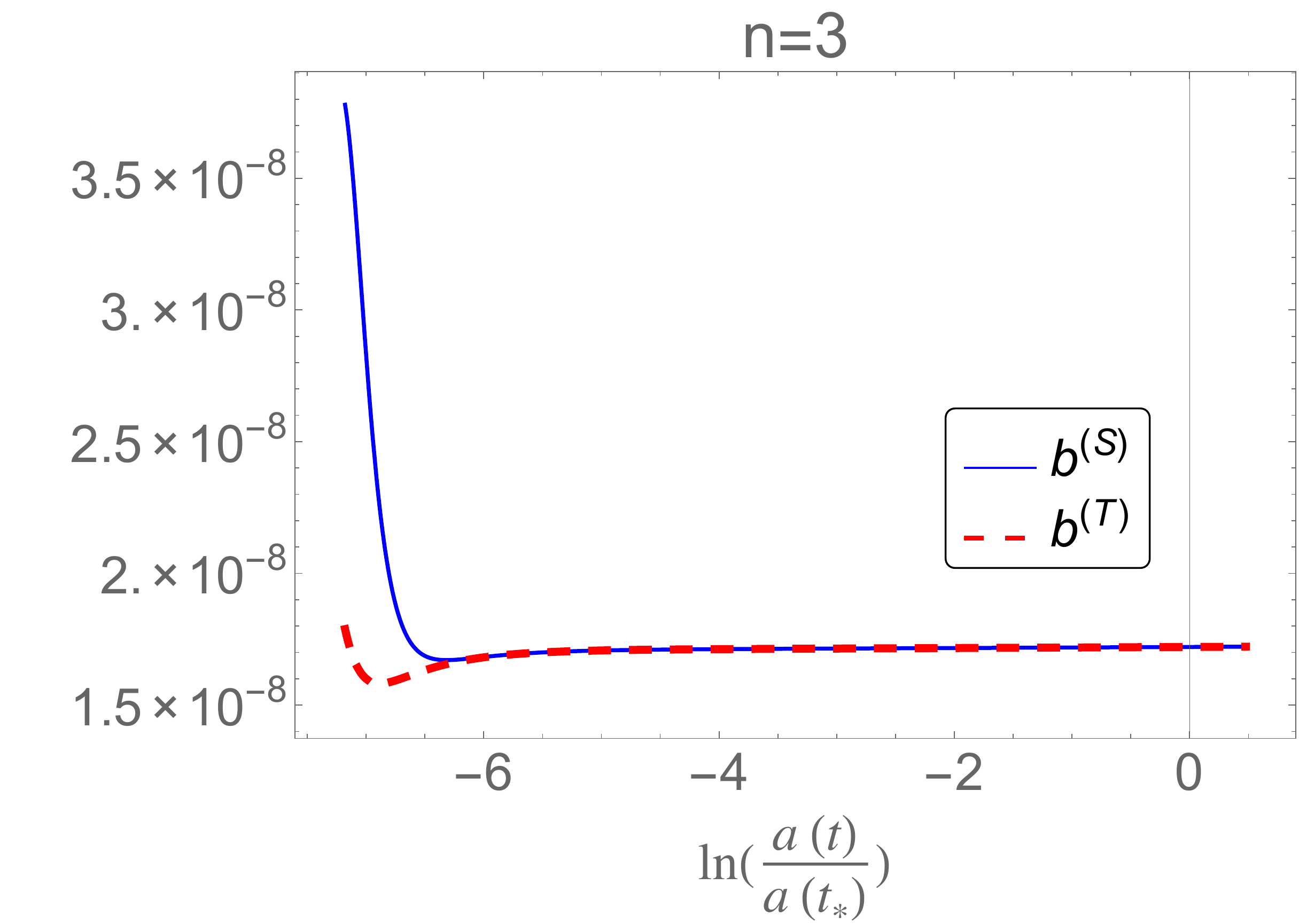}
\ig[width=0.47\textwidth]{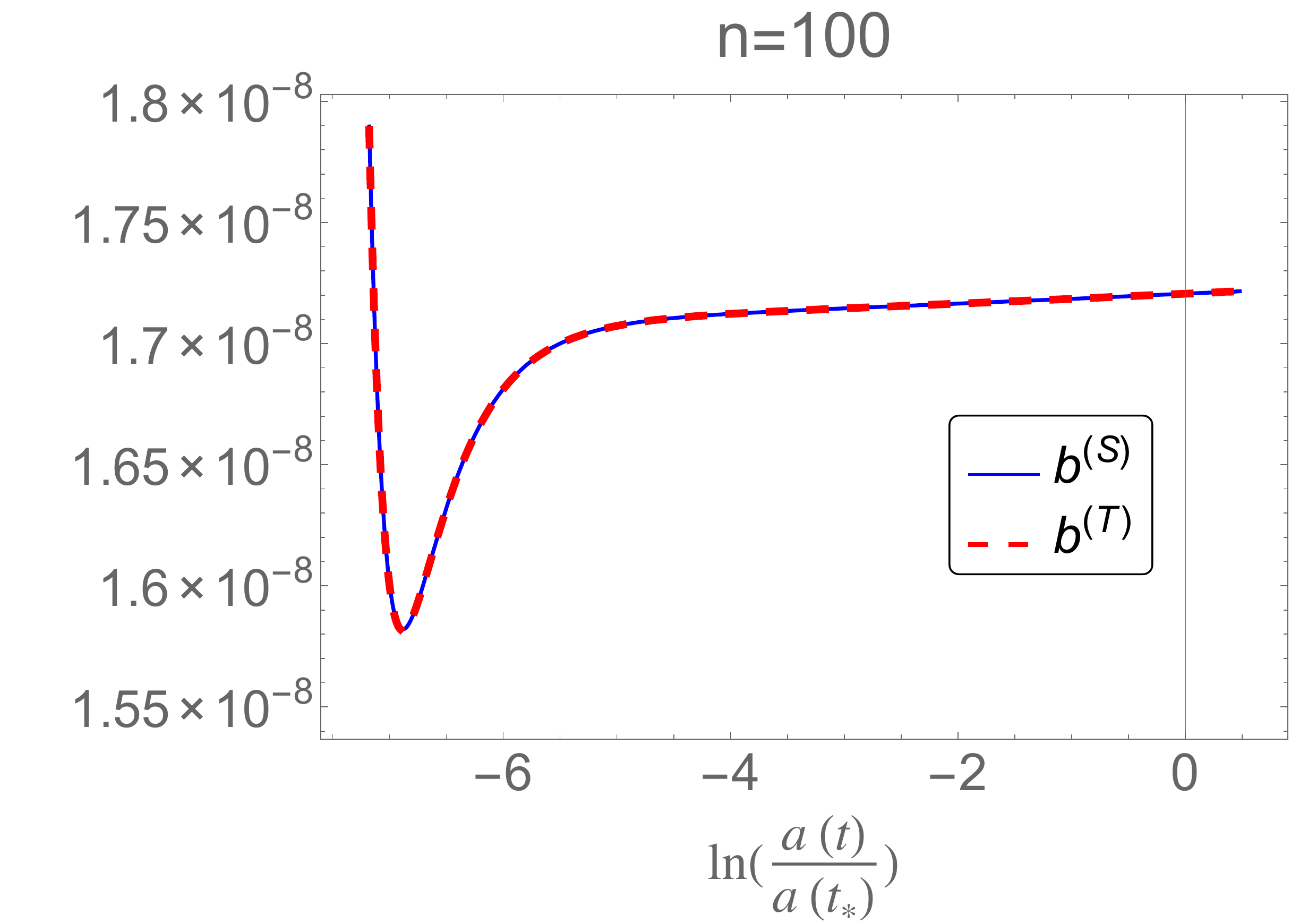}
\caption{Friction terms for scalar and tensor modes plotted for $n=3$ (left panel)
and $n=100$ (right panel) with respect to the number of \efolds from the time when
the pivot modes exits the horizon during slow-roll. It is evident that $b^{(S)}$ and
$b^{(T)}$ differ from each other for $n=3$ at early times while they are practically
the same for $n=100$.}
\label{fig:bterms}
\efig
\bfig
\ig[width=0.475\textwidth]{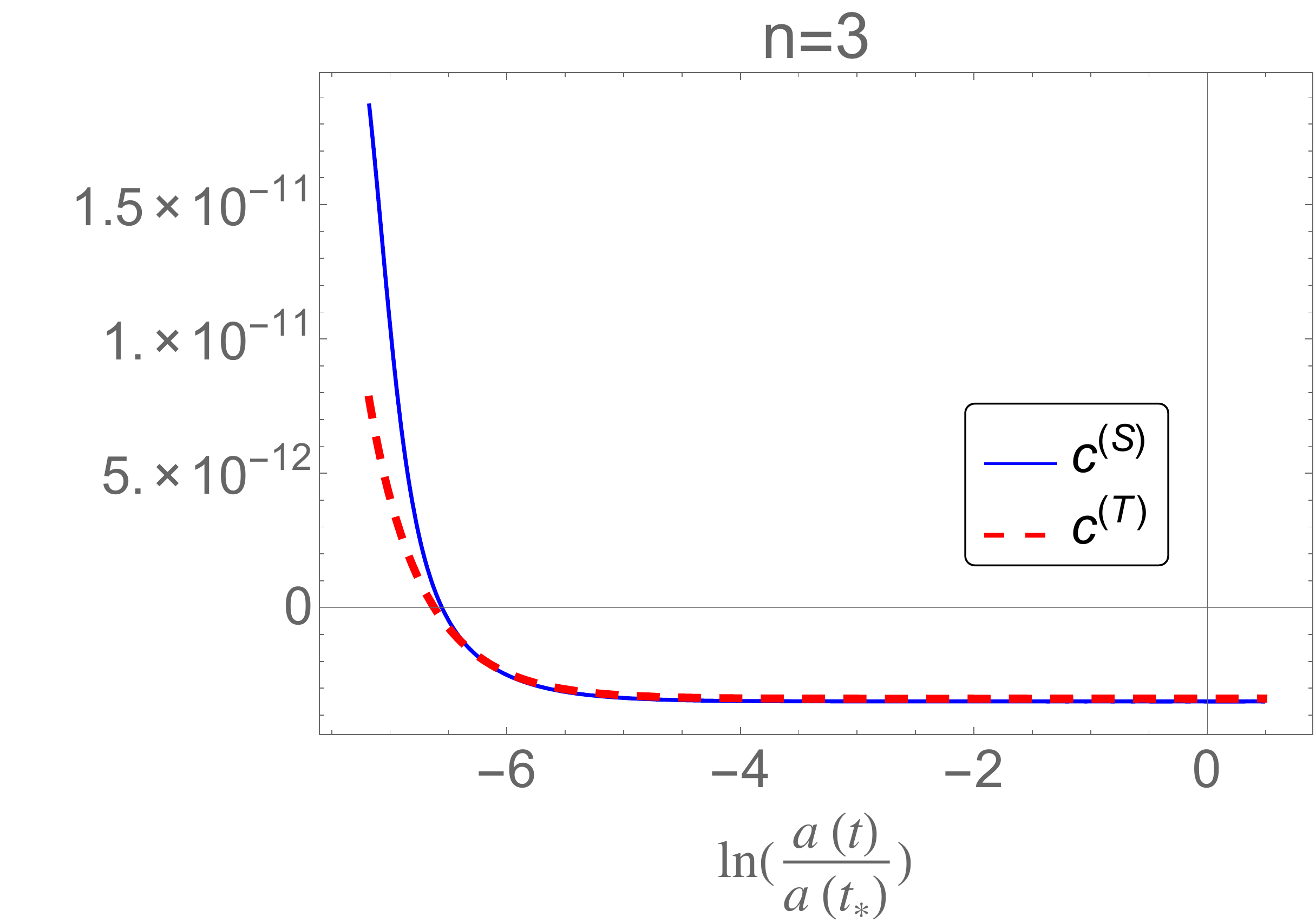}
\ig[width=0.47\textwidth]{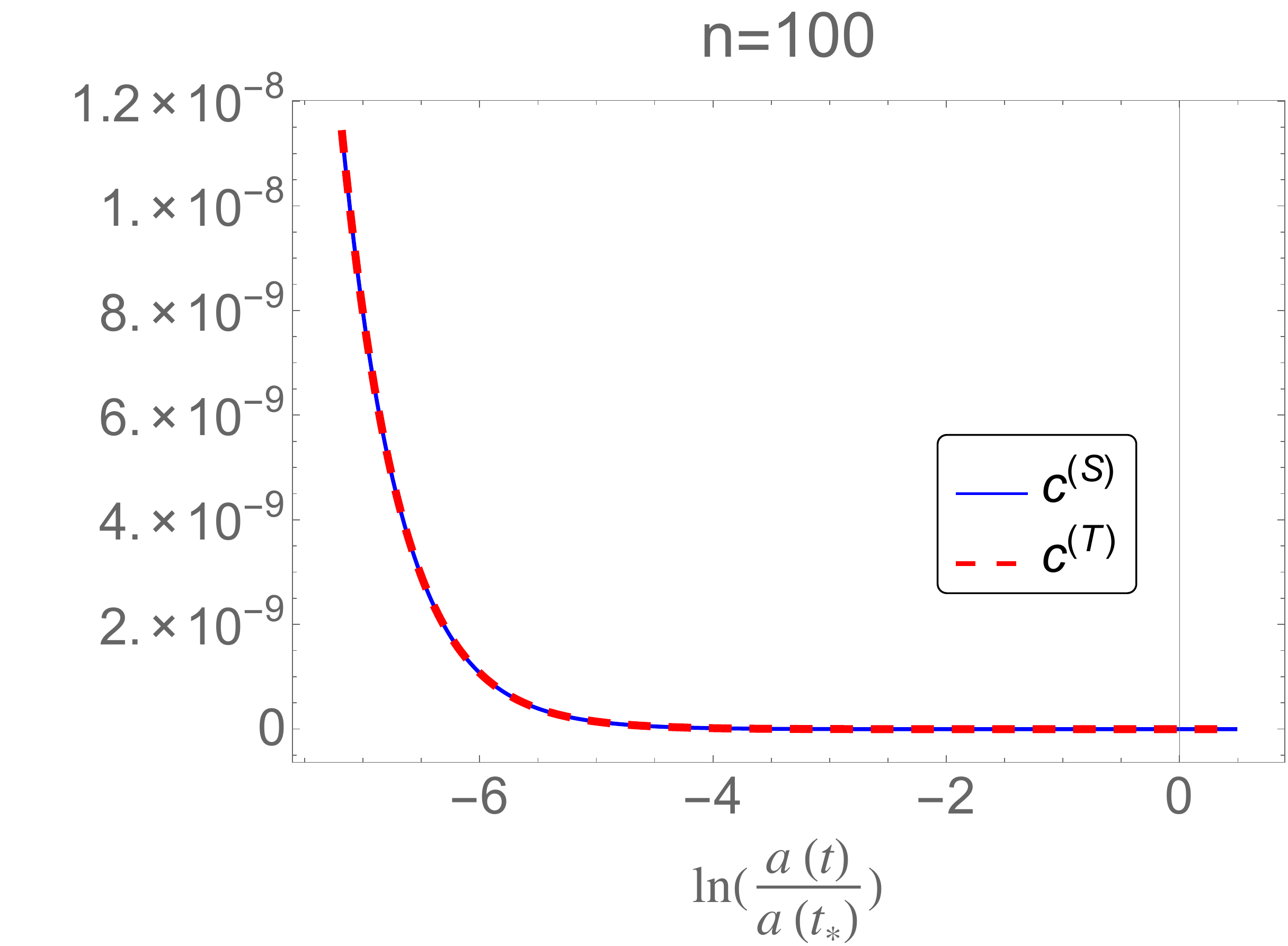}
\caption{Potential terms for scalar and tensor modes plotted for $n=3$ (left panel)
and $n=100$ (right panel) with respect to the number of \efolds from the time when
the pivot modes exits the horizon during slow-roll. As was the case for the friction
terms in \fref{fig:bterms}, $c^{(S)}$ and
$c^{(T)}$ differ from each other for $n=3$ at early times while they are practically
the same for $n=100$.}
\label{fig:cterms}
\efig

\bfig
\ig[width=0.47\textwidth]{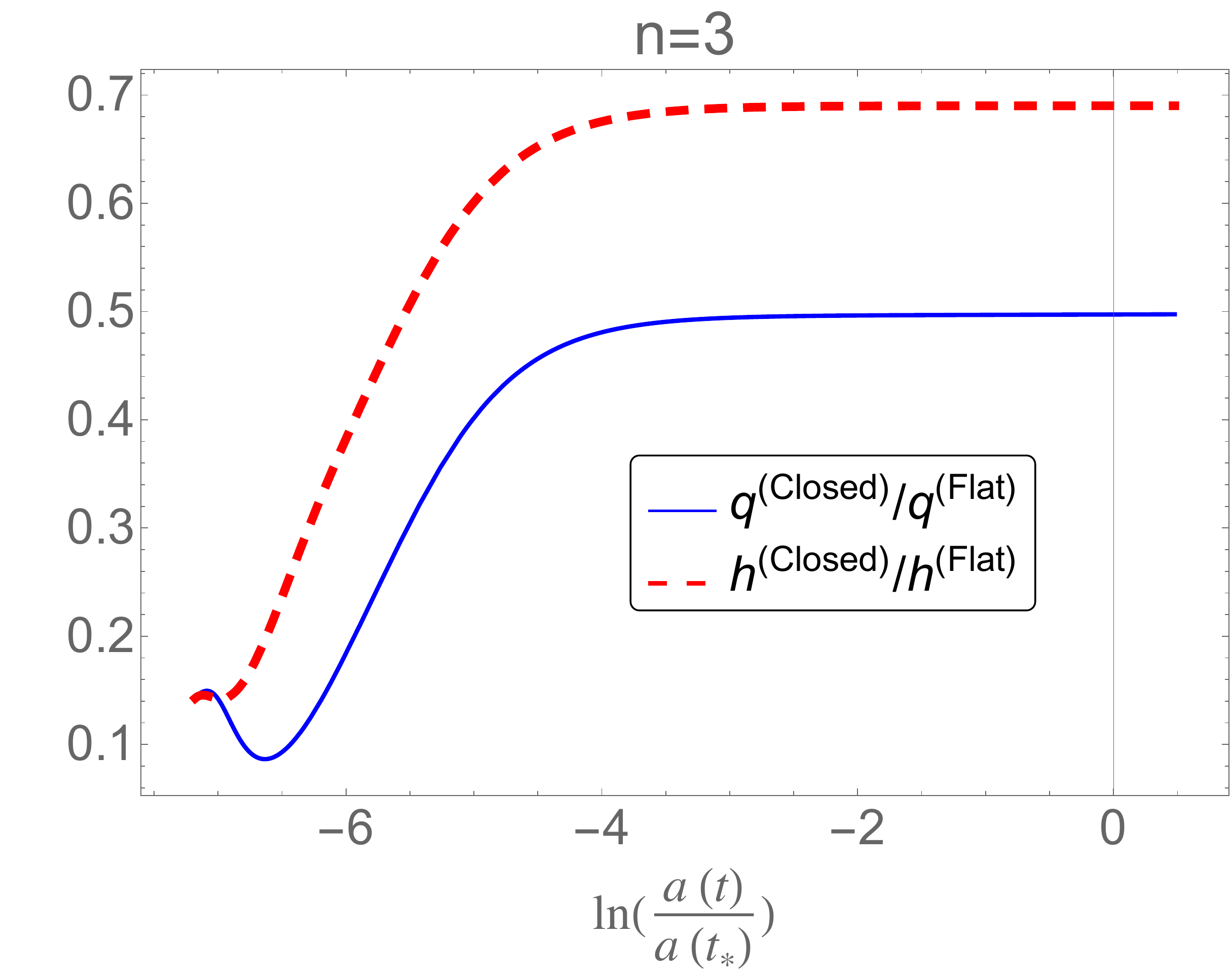}
\ig[width=0.47\textwidth]{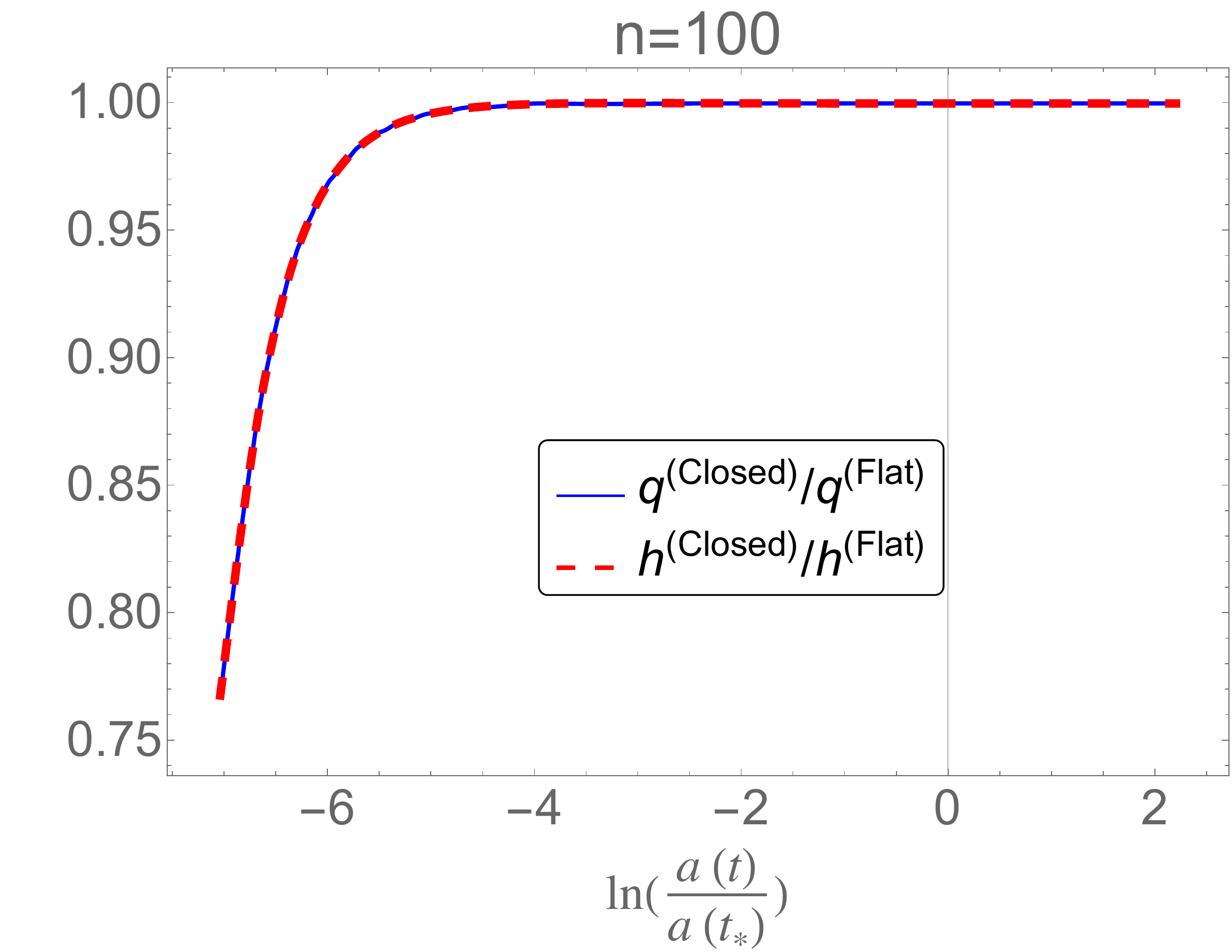}
\caption{Evolution of scalar and tensor modes for spatially closed FLRW model with
$\omk=-0.005$. For the clarity of presentation we have plotted $q=v/a$ and $h=\mu/a$. 
It is evident that the two modes evolve differently for $n=3$ but
in a very similar way for $n=100$. In particular the amplitude of the scalar
modes is lower than that of the tensor modes for $n=3$ after they have exited the
horizon during inflation.}
\label{fig:modes}
\efig

These equations are similar to that of a damped harmonic oscillator where $b^{(S)}$ and
$b^{(T)}$ act like friction and $c^{(S)}$ and $c^{(T)}$ play the role of time
dependent potential. Figures \ref{fig:bterms} and \ref{fig:cterms} respectively show the
evolution of the friction and potential terms for scalar and tensor perturbations. In
both figures the left panels correspond to the longest wavelength mode with $n=3$ and
the right panels correspond to the mode with $n=100$. It is evident that both
friction and potential terms are different for the tensor and scalar modes for
$n=3$, while they are practially the same for the scalar and tensor modes 
for $n=100$. That is, for small $n$, the presence of spatial curvature affects the 
evolution of tensor and scalar modes quantiatively differently. For large $n$,
however, since both the friction and potential terms for scalar and tensor modes are 
indistinguishable from each other the evolution of tensor and scalar modes is 
very similar. This behavior is apparent in \fref{fig:modes}. The scalar
mode with $n=3$ evolves differently from the tensor one and has a lower amplitude
after the horizon exit. Whereas, for $n=100$ both the scalar and the tensor modes
have the same amplitude. This indicates that in a spatially closed FLRW model, there
is a scale dependent correction to the power spectra of the tensor and scalar
perturbations. 

\bfig
  \includegraphics[width=0.47\textwidth]{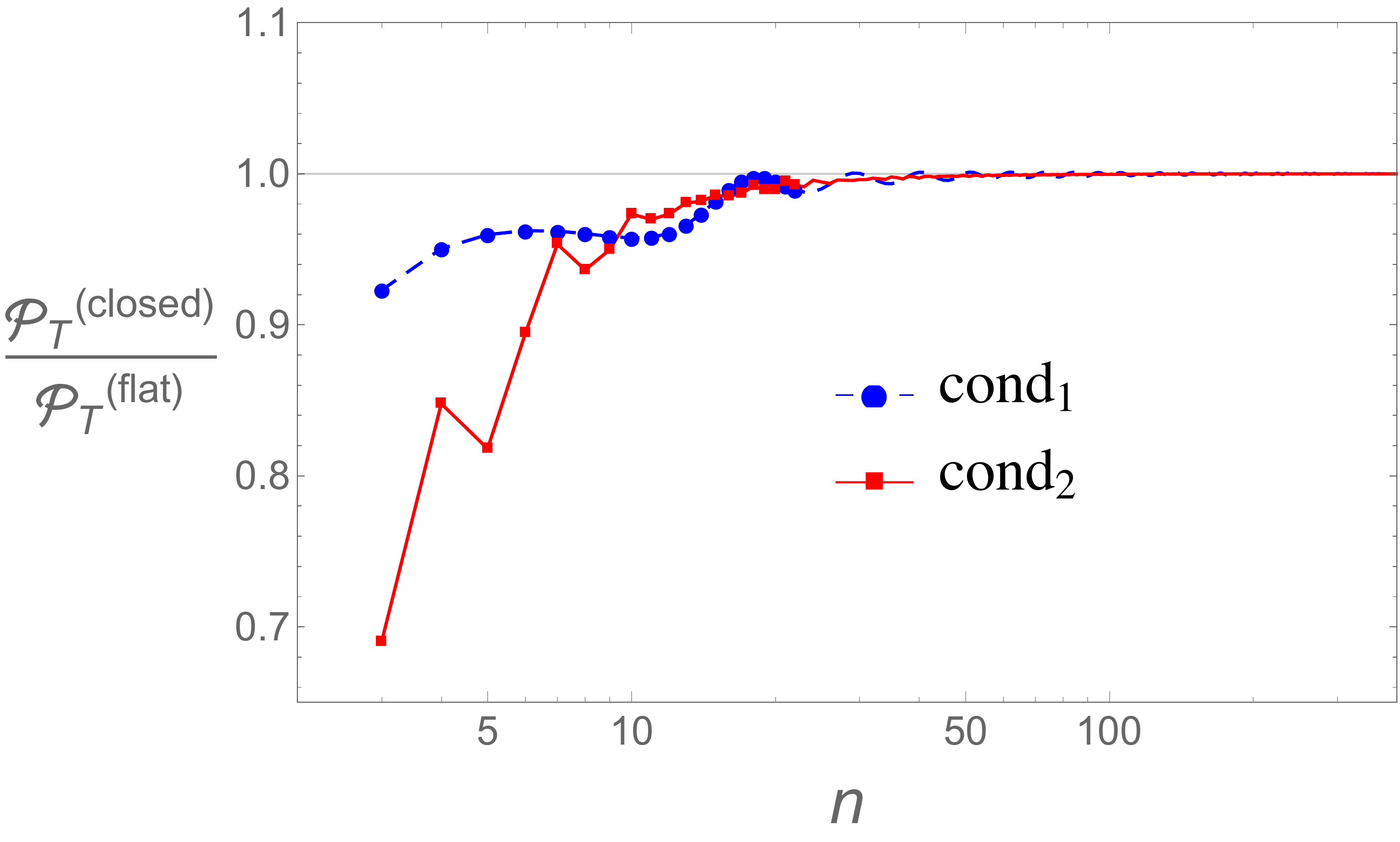}
\hskip0.5cm
  \includegraphics[width=0.47\textwidth]{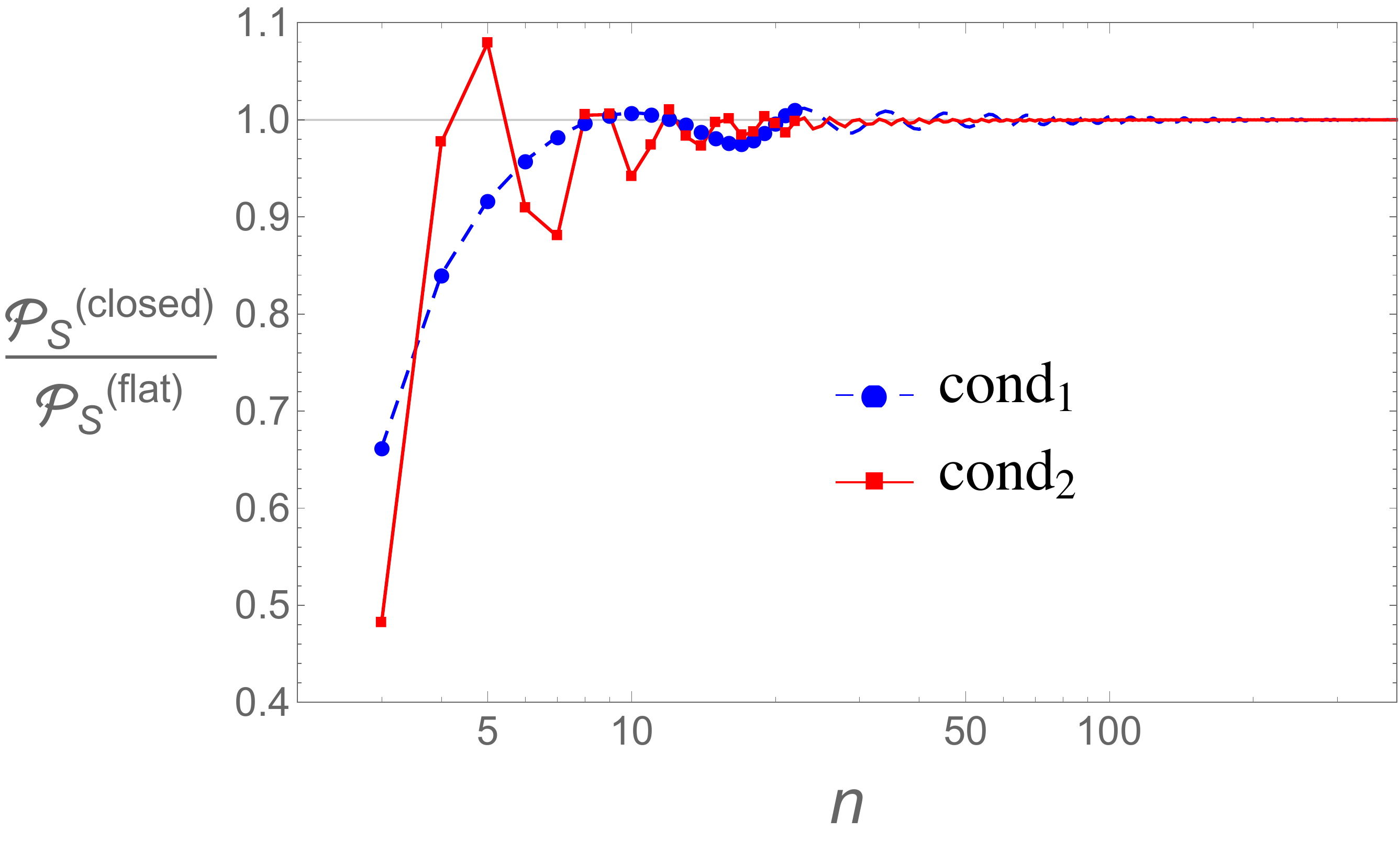}
  \caption{{\it Left panel:} Ratio of the tensor power spectrum in the closed FLRW
model with $\omk=-0.005$ to that in the flat model. 
{\it Right panel:} Ratio of the scalar power spectrum
in the closed FLRW model with $\omk=-0.005$ to that in the flat model. 
Both spectra show suppression of power at small $n$. However, the scalar modes are
significantly more suppressed in the closed Universe as compared to the tensor modes. 
The two curves in each figure correspond to two different initial conditions for the
background geometry: `${\rm cond}_1$' corresponds to \eref{eq:staroini} 
(matching with the best fit values for $\as$ and $\ns$ at $t_\star$) and
`${\rm cond}_2$' corresponds to \eref{eq:staroiniextreme} (generating maximal
suppression). In both figures the spectra are discrete in $n$ (for clarity
discrete points are shown only for $n<20$).  
}
\label{fig:primpowerspectrum}
\efig

The left panel of \fref{fig:primpowerspectrum} shows the ratio of the tensor power
spectrum at the end of inflation in the closed FLRW model with $\omk=-0.005$ to that 
in the flat FLRW model with $\omk=0$. The ratio approaches unity for modes with large $n$ describing the
short wavelength modes which exit the curvature radius later
during inflation. At these times the effect of spatial curvature on the evolution of 
both the background and perturbations is negligible and the background spacetime can 
be very well approximated by a quasi-de Sitter spacetime. For small $n$, which 
correspond to long wavelength modes, however, the ratio is smaller than unity.
That is, the tensor power spectrum is suppressed in the spatially closed model compared 
to the spatially flat model for these modes. As the figure shows, there is as much as
5-30\% suppression in power for modes with $n<10$. The scalar power spectrum (shown
in the right panel of \fref{fig:primpowerspectrum}), for the spatially closed FLRW
model also shows suppression for modes with small $n$ and approaches that for the
flat model as $n$ increases. For the same inflationary background geometry, the
suppression in the scalar power spectrum is larger and is as large as 15\% for $n=10$ and
50\% for $n=3$. 

Hence, the presence of spatial curvature modifies the background geometry and the 
evolution of the quantum metric perturbations in such a way that the power is 
suppressed for \textit{both} the tensor and scalar models at long wavelength modes. 
However, there are differences in the percentage suppressions in tensor and scalar
modes. This is quite different from power suppression mechanisms proposed in the
spatially flat FLRW model by either envisaging a fast roll phase prior to the onset of the usual slow-roll phase
\cite{ContaldiFastRoll,ClineFastRoll,JainFastRoll, PedroFastRoll, LelloFastRoll} 
or by considering 
excited initial states which may arise due to some pre-inflationary physics
\cite{aan3}. 
In those proposals, the relative change to both power spectra (tensor and scalar) are 
very similar to each other, unlike what we find in \fref{fig:primpowerspectrum} for the 
spatially closed FLRW model. 

\bfig
 \begin{center}
 \includegraphics[width=0.7\textwidth]{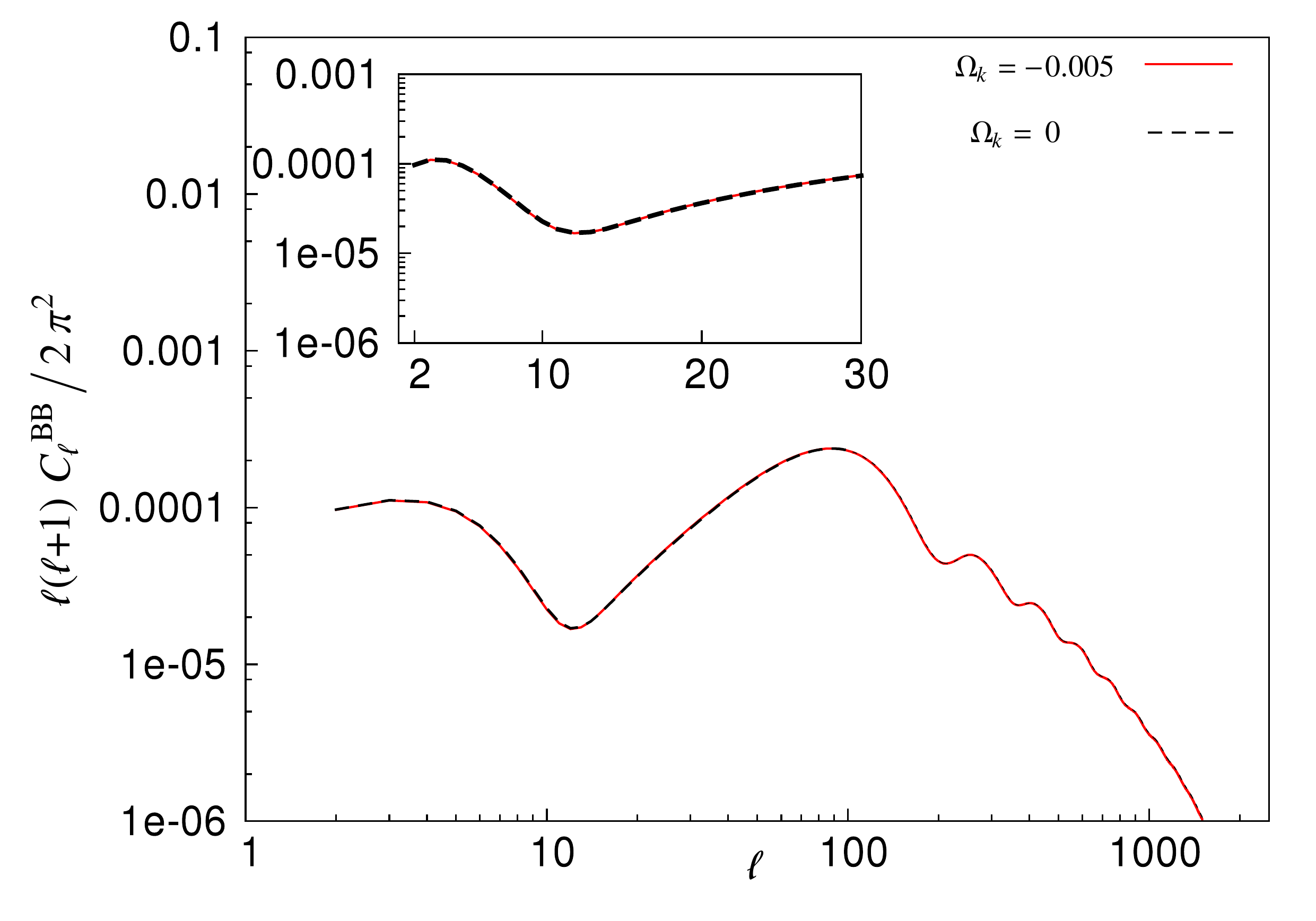}
 \end{center}
 \caption{The B-mode polarization spectrum $\cellbb$ in the CMB for $\omk=-0.005$
(solid red curve) and $\omk=0$ (black dashed curve). In spite of the differences in
the tensor primordial power spectrum (\fref{fig:primpowerspectrum}) at low $n$, the
observable $\cellbb$ remains practically unchanged in the presence of the spatial
curvature for $\omk=-0.005$ with initial conditions corresponding to 
${\rm cond}_2$ for which the wavelength of the largest 
observable mode is comparable to the Hubble horizon at the onset of inflation.}
 \label{fig:cBB}
\efig
Let us now discuss how the CMB B-mode polarization spectrum is affected due to the
modifications in the tensor primordial spectrum shown in
\fref{fig:primpowerspectrum}. We use the publicly available Boltzmann solver
\texttt{CAMB} \cite{Lewis:1999bs,camb_notes} to evolve the primordial power spectrum until the surface
of last scattering and compute the resulting $\cellbb$ shown in \fref{fig:cBB}. As
the figure shows, the resulting $\cellbb$ for $\omk=-0.005$ is practically indistinguishable
from that in the spatially flat model (with $\omk=0$). This happens because the
corrections to the tensor primordial power spectrum is limited to very low
wavenumbers ($n<15$), which correspond to super-horizon modes. This is true not only for 
initial conditions corresponding to the best fit values of $\as$ and $\ns$ 
(${\rm cond}_1$), but also
for the initial conditions that generate maximal suppression of power in the primordial
tensor power spectrum (${\rm cond}_2$). Thus, this result is robust under changing the
initial conditions for the background.
\vskip0.5cm
\noindent {\it Remark:} If one considers the coupling between long and short wavelength 
modes along the lines of
\cite{Erickcek:2008jp,Erickcek:2008sm,Schmidt:2012ky,Agullo:2015aba,Adhikari:2015yya}, 
the corrections to the primordial power spectrum at small $n$ could influence
the computation of super-horizon modulation to the observed two-point function for
both tensor and scalar modes depending on the type of coupling considered.

\subsection{Tensor-to-scalar ratio and the slow-roll consistency relation}
\label{s3.2}
\bfig
 \begin{center}
 \includegraphics[width=0.6\textwidth]{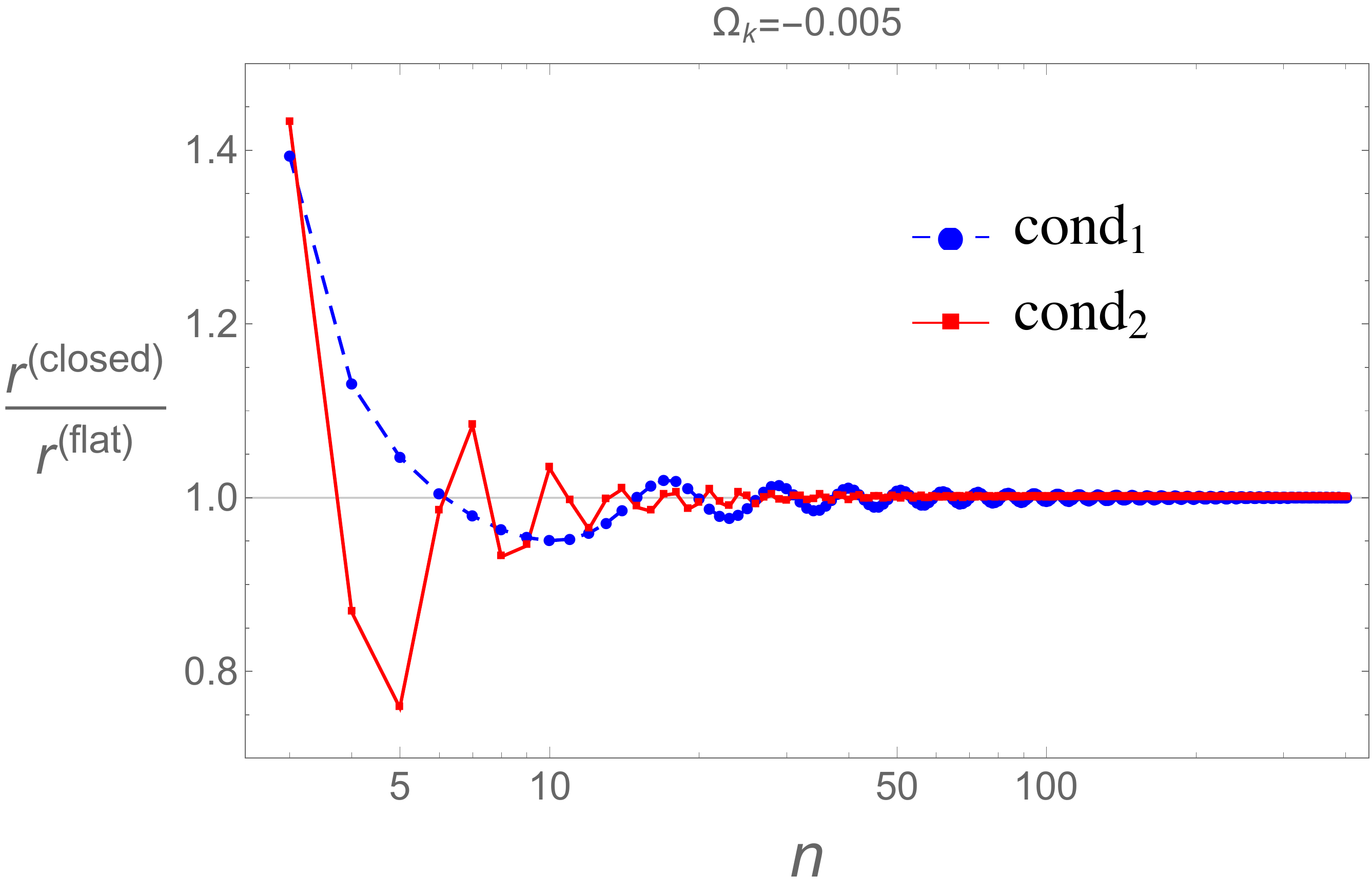}
 \end{center}
 \caption{The tensor-to-scalar ratio $r$ in the spatially closed model compared with
that in the flat model. For small $n$, which correspond to modes exiting the Hubble
horizon at early times during inflation, $r$ in the closed model is modified compared
to the flat model. For large $n$ modes, which exit the Hubble horizon later during
inflation, $r$ in the closed model is practically the same as in the flat model.
As in \fref{fig:primpowerspectrum}, ${\rm cond}_1$ and ${\rm cond}_2$
correspond to the background initial conditions \eref{eq:staroini} and
(\ref{eq:staroiniextreme}) respectively.}
 \label{fig:tensortoscalar}
\efig
\bfig
 \begin{center}
 \includegraphics[width=0.6\textwidth]{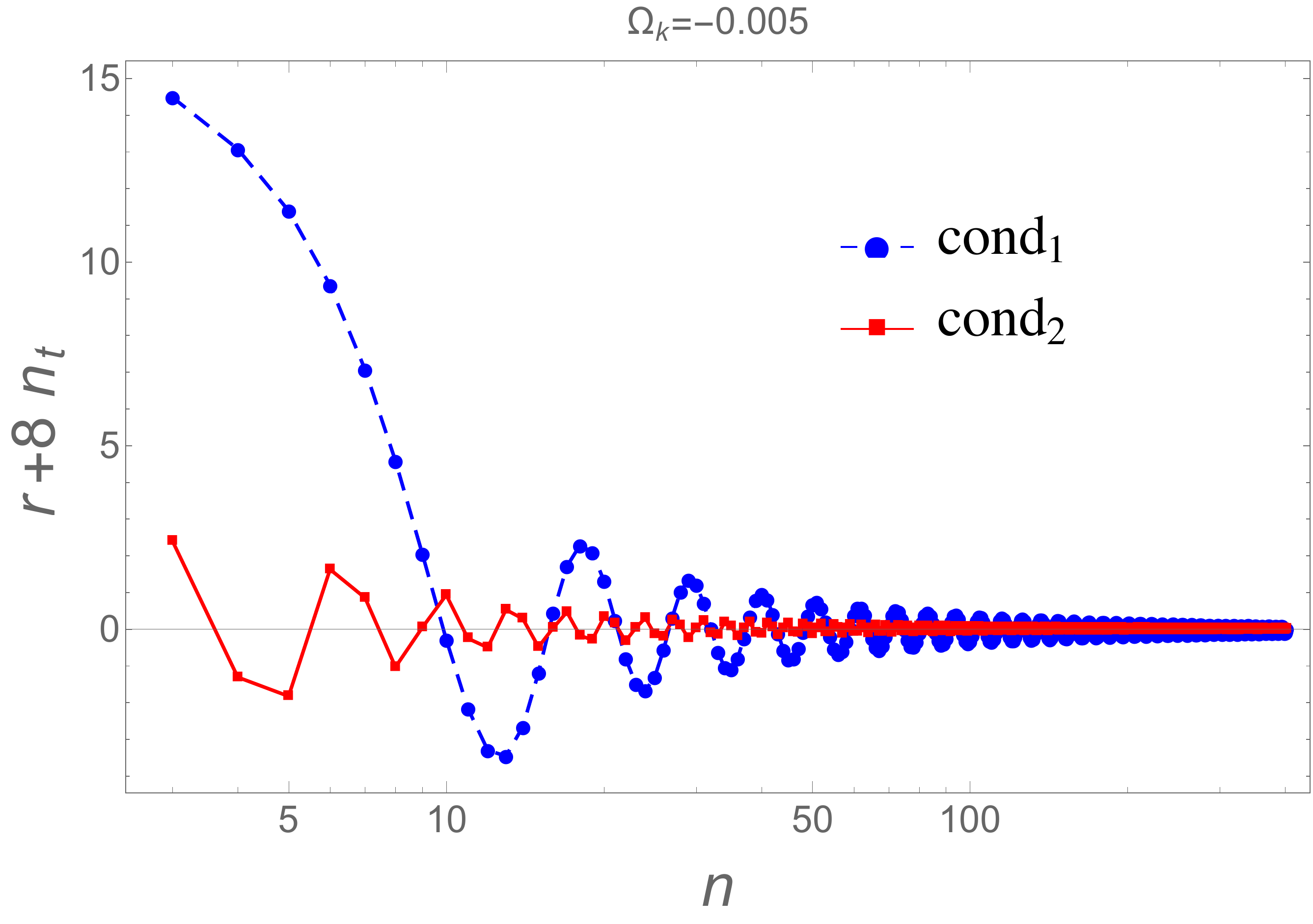}
 \end{center}
 \caption{The deviation from the consistency relation ($r+8\nt=0$) for $\omk=-0.005$
plotted with respect to the spherical wavenumber $n$. It is evident that for small $n$ 
deviation from $r+8\nt=0$ is more prominent. These deviations approach zero as $n$
increases. As before, ${\rm cond}_1$ and ${\rm cond}_2$ respectively
correspond to the background initial conditions \eref{eq:staroini} and
(\ref{eq:staroiniextreme}) respectively.}
 \label{fig:consistency}
\efig
As shown in \fref{fig:primpowerspectrum}, the modifications to the tensor and scalar
power spectrum for a spatially closed model are different quantitatively. Therefore, the
tensor-to-scalar ratio $r$ also shows deviation from the predictions of the standard
inflationary scenario for a flat FLRW model. \Fref{fig:tensortoscalar} shows
the ratio of $r$ in the closed FLRW model with $\omk=-0.005$ to that in the flat FLRW
model.
In each figure the dashed (blue) and the solid (red) curves correspond to initial
conditions given in \eref{eq:staroini} and \eref{eq:staroiniextreme} respectively. As
discussed previously in this section, the effect of the spatial curvature on the
evolution of the modes with large $n$ is minimal. Therefore, the ratio $r^{\rm
(closed)}/r^{\rm (flat)}$ approaches unity for large $n$. However,
both the scalar and tensor power spectra are modified and as a result the
tensor-to-scalar ratio for the closed model shows deviation from the flat model for
small $n$. The differences are most prominent for modes with $n<10$ which correspond
to super horizon modes for $\omk=-0.005$. So, as far as the observable modes are
concerned, the correction to $r$ due to the spatial curvature is small.\footnote{The
violation of the consistency relation has also been obtained due to the
pre-inflationary physics, see e.g. \cite{aan3,Ashoorioon:2005ep}.} 

Let us now understand the impact of spatial curvature on the slow-roll consistency
relation. One of the important features of the standard slow-roll inflation in a flat
FLRW model, assuming Bunch-Davies initial conditions for the perturbations, is the 
relation between $r$ and tensor spectral index $\nt$:  $r^{\rm (flat)}\approx-8\nt$,
where `$\approx$' indicates that the relation is valid only in the slow-roll approximation. 
This relation is known as the slow-roll consistency relation and it holds regardless of the 
choice of potential and hence can be used as test of the standard inflationary scenario. 
In the scenario considered here, $r$ acquires scale dependent modifications due to the 
non-zero spatial curvature (see \fref{fig:tensortoscalar}). So, one is naturally led
to ask: Is the consistency relation also modified? \Fref{fig:consistency} shows the
behavior of $r+8\nt$ (which is zero for the standard scenario) in the presence of 
spatial curvature.\footnote{In order to compute the tensor spectral index in $\s3$ 
spatial topology we use the following definition of $\nt$: 
$\nt = \frac{n^2-3}{n} \frac{d \ln \Pt}{d n}$.} It is evident from the figure that at 
large $n$ the consistency relation holds just as in the standard case, while for small 
$n$ there are deviations for the closed model.

\section{Summary and outlook}
\label{s4}
One of the attractive features of the inflationary scenario is that it `dilutes' away
the information about the spatial curvature. The CMB observations confirm this by putting 
strong constraints on the spatial curvature ($\omk$) today. During inflation, the 
contribution from the spatial curvature to the total curvature of the spacetime goes as 
$a^{-2}$ while that from the scalar field remains almost constant. Therefore, even if the 
spatial curvature of the spacetime was important before the onset of inflation, it becomes
negligible with respect to the energy density of the matter field within a few
e-folds. As a result, the spacetime is extremely well approximated by a flat FLRW
model at the end of inflation and in the post-inflationary era. This suggests that
while most of the inflationary and post-inflationary evolution of the spacetime
remains unaffected by the spatial curvature, the early phases of inflation can be
modified by a non-vanishing $\omk$. In that case, long wavelength modes which 
exit the Hubble horizon at these early times can carry imprints of spatial curvature in 
their power spectrum. In a recent paper \cite{bgy1}, we considered the evolution of 
gauge-invariant scalar perturbations in a spatially closed FLRW model and found that the 
primordial scalar power spectrum is indeed modified leading to suppression of power 
at large angular scales in the CMB temperature anisotropy spectrum $\celltt$.

In this paper, we extended the analysis to include tensor modes. Similarly to the
scalar perturbations, the tensor power spectrum is modified due to the presence of
closed spatial curvature for modes with small $n$ that exit the curvature radius at
early phases of inflation. For these modes, the power in the primordial tensor spectrum is
suppressed compared to that in the flat inflationary FLRW spacetime. However, we find
that the relative suppression for tensor modes is smaller than that for the scalar
modes. 
The suppression in the tensor power spectrum is weak
and limited to very small $n$, which correspond to super-horizon modes. Consequently,
the resulting observable polarization anisotropy spectrum $\cellbb$ in closed 
FLRW model differs less than a percent from the flat FLRW model. Since the tensor and scalar spectra are modified differently by 
the spatial curvature, the tensor-to-scalar ratio at the long wavelength modes acquires 
scale dependent corrections which further leads to the violation of the standard slow-roll
consistency relation obtained for the flat FLRW model. Hence, although the
modifications in the tensor spectrum due to closed spatial curvature do not have a
direct observable imprint in the B-mode polarization signal, they lead to violation
of the standard slow-roll consistency relation for long wavelength modes. This deviation 
from the standard relation, if observed, can be used to further refine the constraints on 
spatial curvature of the Universe.

Since the effects of spatial curvature is more dominant in the early phases of
inflation, it is expected that the pre-inflationary dynamics of the spacetime and its
Planck scale behavior in the presence of closed spatial curvature will be 
different from the spatially flat model. A priori, it is not clear if inflation will
even happen if one provides initial conditions in the deep Planck regime. These
issues have been investigated in the setting of loop quantum cosmology (LQC) for the
spatially flat FLRW model
\cite{Ashtekar:2011rm,Ashtekar:2009mm,Corichi:2010zp,aan3,bg1,bg2}. It is expected
that the interplay between the quantum gravity corrections due to LQC and the spatial
curvature may bring out interesting features that could be relevant for observations. 
In future work \cite{bgy3}, we will study the Planck dynamics of the closed FLRW spacetime 
using the LQC framework developed in \cite{apsv}, and study the evolution of quantum
perturbations on the quantum modified background geometry.

We considered spacetimes with spatial sections isometric to the three-sphere
$\mathbb{S}^3$, the simplest global geometry compatible with a constant positive
spatial curvature. The local geometry constrains the global topology but does not fix
it uniquely. 
In this work, we focused on the
analysis of geometric effects due to the presence of spatial curvature, but our
framework can also be applied to the study of observational signatures of the spatial
topology in locally spherically Universes. In particular, solutions for the equations
of motion for tensor perturbations on nontrivial topologies can always be written as
linear superpositions of normal modes on $\mathbb{S}^3$, allowing the evolution of
perturbations and the primordial power spectrum to be determined for arbitrary
topologies from the results herein discussed for the dynamics of the perturbations on
$\mathbb{S}^3$. 

\vskip0.04cm
\section{Acknowledgements}{We are grateful to Abhay Ashtekar for discussions.
BB would also like to thank Steve Carlip
and Glenn Starkman for discussions on the topology of the Universe. 
This work was supported by NSF grant PHY-1505411 and the Eberly research funds of 
Penn State. BB is grateful for receiving a Mebus Fellowship and NY acknowledges support 
from  CNPq, Brazil. This work used the Extreme Science and Engineering Discovery
Environment (XSEDE), which is supported by National Science Foundation grant number
ACI-1053575.}
\vskip1cm


\end{document}